%% file: NuPhys2017_AD_v3.tex
\documentclass[12pt]{article}
\usepackage{graphicx}

\newcommand\pubnumber{SNSN-323-63}
\newcommand\pubdate{\today}

\def\krakow{Institute of Nuclear Physics Polish Academy of Sciences,\\
 PL-31342 Krakow, Poland}

\def\Title#1{\begin{center} {\Large #1 } \end{center}}
\def\Author#1{\begin{center}{ \sc #1} \end{center}}
\def\Address#1{\begin{center}{ \it #1} \end{center}}

\newcommand\pubblock{\rightline{\begin{tabular}{l} \pubnumber\\
         \pubdate  \end{tabular}}}
\newenvironment{Abstract}{\begin{quotation}  }{\end{quotation}}
\newenvironment{Presented}{\begin{quotation} \begin{center} 
             PRESENTED AT\end{center}\bigskip 
      \begin{center}\begin{large}}{\end{large}\end{center} \end{quotation}}
\def\Acknowledgements{\bigskip  \bigskip \begin{center} \begin{large}
             \bf ACKNOWLEDGEMENTS \end{large}\end{center}}

\input econfmacros.tex

\begin{document}
\begin{titlepage}
\pubblock

\vfill
\Title{Study of the neutral current interactions with production 
of neutral pions  in the T2K off-axis near detector}
\vfill
\Author{ Anna Dabrowska\\
for the T2K Collaboration}
\Address{\krakow}
\vfill
\begin{Abstract}
Abstract\\This analysis aims to study inclusive neutrino-induced neutral
current (NC) interactions that produce at least one neutral pion ($\pi^{0}$) in the
final state (NC$\pi^{0}$) in the T2K off-axis near detector ND280.

The motivation for this study is to better understand/eliminate the
background to electron neutrino appearance analyzed at the far
detector, Super-Kamiokande. As $\pi^{0}$ decays are background to the
appearance search, a good knowledge of both inclusive and exclusive
neutrino interaction cross sections is key to improving associated
systematic uncertainties in T2K oscillation analyses.

Results from this study can be compared to a number of Monte Carlo
samples produced using different neutrino interaction generators.
This can yield information about models of neutrino-nucleon
interactions. Further comparisons can be made with the inclusive CC$\pi^{0}$ sample already
obtained from  a T2K near detector analysis.

In this poster, the selection criteria used to collect inclusive neutral current 
neutrino interactions on a plastic scintillator (CH) with at least one $\pi^{0}$ in the final state
will be presented, including a discussion of how the selection criteria were improved.
I will also  discuss the changes in the selections, including the use of a fitting method
to improve measurement of the NC$\pi^{0}$ interaction rate.

\end{Abstract}
\vfill
\begin{Presented}
Conference NuPhys2017: Prospects in Neutrino Physics\\
London, UK,  December 20--22, 2017
\end{Presented}
\vfill
\end{titlepage}
\def\thefootnote{\fnsymbol{footnote}}
\setcounter{footnote}{0}

\section{T2K experiment}

T2K, a long-baseline neutrino experiment~\cite{Abe}, is located in  Japan. 
A high-intensity beam of muon neutrinos/anti-neutrinos produced at the J-PARC accelerator complex 
is sent towards the near detector facility (the ND280 and INGRID detectors, located 280 m away from the neutrino source) and the far water Cherenkov detector, Super-Kamiokande (295 km away). 
The beam is produced by the conventional method. 
The protons accelerated to 30 GeV hit a graphite target, producing hadrons including pions
and kaons. Charged hadrons are then focused and sent to a
decay tunnel where the pions and kaons decay in flight, producing neutrinos (or anti-neutrinos  by the reversing  current in magnetic horns). 
The main purpose of  the T2K experiment is the measurement of the neutrino oscillation parameters based on the  comparison of the intensity and the composition of the neutrino beam between the near and far detectors.
The T2K experiment is also capable of providing information on neutrino-nucleus cross sections at energies around 1 GeV, thanks to a large amount of target material present in the near detector facility.
The ND280, shown in Figure~\ref{fig:01} (left), is used to constrain flux and cross-section systematics for oscillation analysis; it measures the flux and cross section before the oscillations occur. 
It consists of several sub-detectors: the Pi-Zero Detector (P0D) and  Tracker as the inner detectors, both surrounded by the electromagnetic calorimeter (ECAL) and then by the Side Muon Range Detector (SMRD). All detector components, except the SMRD, are placed inside a 0.2 T magnetic field. The P0D subdetector, placed  upstream inside the magnet is a "sandwich" of scintillator planes, lead and brass plates, and a water target.  It is optimized for the measurement of $\pi^{0}$ production.
Gamma rays from a $\pi^{0}$ decay are converted to electromagnetic showers in the  lead plates, and are measured in the scintillator detectors.
Downstream of the P0D, the Tracker consists of two  Fine Grained Detectors (FGDs) separated by three Time Projection Chambers (TPCs). The TPCs can measure the momenta of charged particles from the curvature of the  tracks in the magnetic field.  The FGDs consist of scintillator bars. They provide the target material for neutrino interactions. 
Particle identfication is based on momentum reconstruction and measurement of energy loss, $dE/dx$, in TPC's.
 The inner part of the detector is surrounded by an electromagnetic calorimeter (ECAL) which can tag escaping electrons and positrons from $\pi^{0}$ decays. 
The SMRD (scintillator slabs installed in the magnet yoke) detector is used to detect  muons.

\section{Motivation}

This analysis aims to study inclusive neutrino-induced ($\nu$) neutral
current (NC) interactions that produce at least one neutral pion ($\pi^{0}$) in the
final state (NC$\pi^{0}$) in the T2K off-axis near detector ND280.
In asymmetrical conversions of a photon derived from the disintegration of $\pi^{0}$, low energy particles can be invisible in the detector, so the detector will only register higher energy electrons or positrons.
As $\pi^{0}$ decays can be the background to the $\nu_{e}$
appearance search, a good knowledge of both inclusive and exclusive
neutrino interaction cross sections are key to improving associated
systematic uncertainties in T2K oscillation analyses.
Such an analysis can help to better understand/eliminate the NC$\pi^{0}$ background to electron neutrino appearance at the far detector Super-Kamiokande.
The results from this study can be compared to a number of Monte Carlo samples produced using different neutrino interaction generators.
This can yield information about models of $\nu$-nucleon interactions.

\section{Selection criteria}

The definition of a signal reaction (NC$\pi^{0}$) is based on particles exiting the nucleus: no charged leptons, at least one  $\pi^{0}$, any number of baryons and mesons ({\textit {X}}), $\nu + {\textit{p}}({\textit{n}}) \rightarrow {\textit{p}}/\pi^{+} + \pi^{0} + \textit{X}$.

\begin{figure}[htb]
\centering
\includegraphics[height=1.5in]{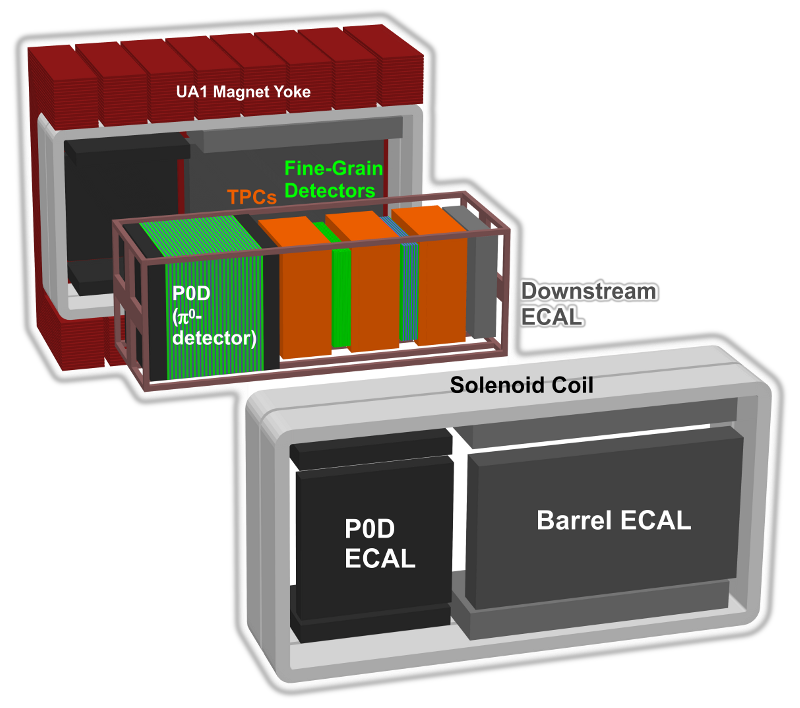}
\includegraphics[height=1.5in]{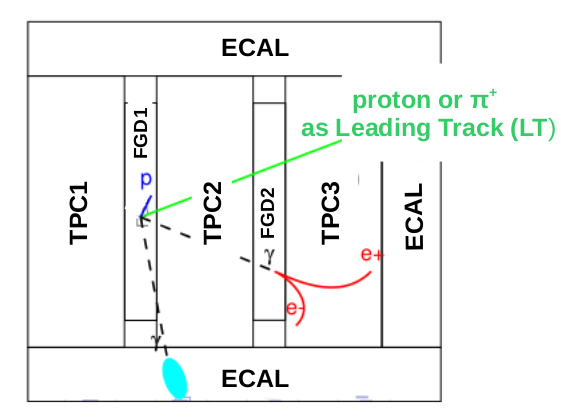}
\caption{The ND280 off axis near detector (left) and a schematic example of a NC$\pi^{0}$ candidate in the ND280 tracker (right)  }
\label{fig:01}
\end{figure}

To identify the NC$\pi^{0}$ event candidate, the position of the $\nu$ interaction vertex, angle of $\gamma$'s (from  $\pi^{0}$) emission  and their energy are reconstructed.
The $\nu$ interaction vertex is selected based on the reconstructed start position of the so-called Leading Track (LT) in the Fiducial Volume (FV) of FGD1.
As the LT the tracks of proton (\textit{p}) or positive pion ($\pi^{+}$) candidates are selected.
The identification of the particle is based on the energy loss in the TPC's.
The momentum of the selected LT should be above 350 MeV/c, to reject electrons and  muons. The likelihood $L_{p, \pi^{+}} > 0.1$ for proton or pion candidates ‬is demanded, where $L$ is defined as:
  \begin{equation}\label{def}
	 L_{\alpha} = L({\alpha}|p_{k}, {Pull}_{\alpha,\beta})  = \frac{\prod_{k}e^{(Pull^{2}_{\alpha,k}|p_{k},\alpha)}}{ \sum_{\beta}\prod_{k}e^{\textrm(Pull^{2}_{\beta,k}|p_{k},\beta)} }
	\end{equation} 
 Based on the energy loss ($dE/dx$) in the k'th TPC as a function of momentum $p_{k}$ for a given particle $\alpha$, the particle identifcation i.e. ${Pull}_{\alpha}$, was calculated:\\
                 
\begin{equation} 
	{Pull}_{\alpha}(p) = \frac{(\frac{dE}{dx})_{expected} - (\frac{dE}{dx})_{measured}}{\sigma}
\end{equation}

where: $(dE/dx)_{measured}$ is the energy deposited per unit length in the
TPC (using a truncated mean algorithm), $(dE/dx)_{expected}$ is the expected
value of the energy deposit for a given particle hypothesis and $\sigma(dE/dx)$
is the expected value of the error.
To reject CC interactions from the selected sample, a muon veto is imposed in the next step of selection.
The muon candidate is defined as a reconstructed track with:
negative charge and ($L_{MIP}  > 0.8$ or $p>500$), and $L_{\mu} > 0.05$, where $L_{MIP}$ = ($L_{\mu}+ L_{\pi})/(1-L_{p})$.
Secondary particles entering FGD1  from neutrino interactions in the upstream part of the ND280 detector can interact  in FGD1. So an event is vetoed if any track is reconstruted in P0D, P0DEcal or TPC1 subdetectors.
Among collected NC interaction candidates, only these with at least one $\pi^{0}$ candidate in the final state are saved in the NC$\pi^{0}$ sample.
The  $\pi^{0}$ identifcation was based on the search for combinations of converting $\gamma$'s i.e.: ECAL showers, $e^{+}$ $e^{-}$ pairs in the FGD/TPC and  
$e^{+}$ or $e^{-}$ tracks in the FGD/TPC (shown in Figure~\ref{fig:01} (right)). For photon conversions in the ECAL we require isolated shower-like objects with electromagnetic energy of more than 50 MeV.
The $e^{+}$, $e^{-}$ tracks in the TPC should be good quality tracks (TPC tracks with at least 18 nodes) with the starting position located in FGD1, $|Pull_{e}| < 2$ and momentum above 50 MeV/c. Additionally, the momentum of positrons should be below 800 MeV/c to reject protons.

\section{Selection results}

\begin{figure}[htb]
\centering
\includegraphics[height=1.5in]{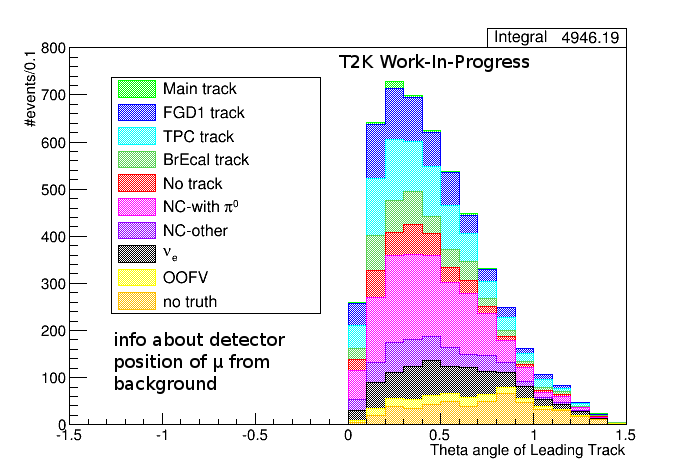}
\includegraphics[height=1.5in]{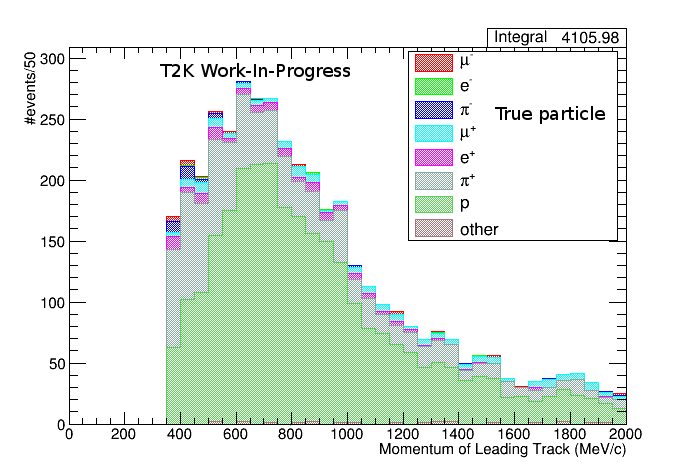}
\includegraphics[height=1.5in]{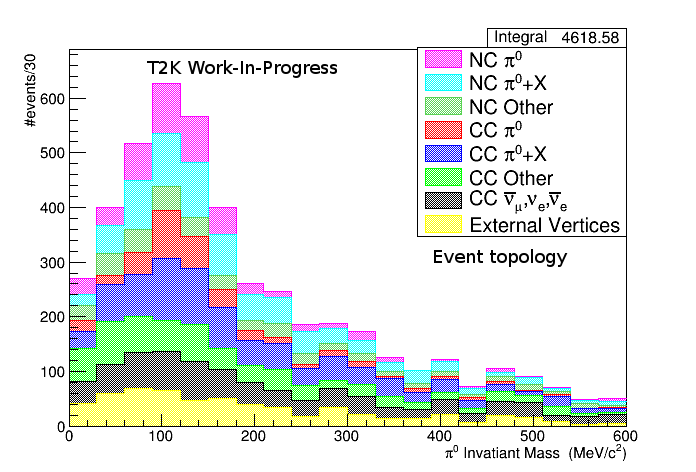}
\includegraphics[height=1.5in]{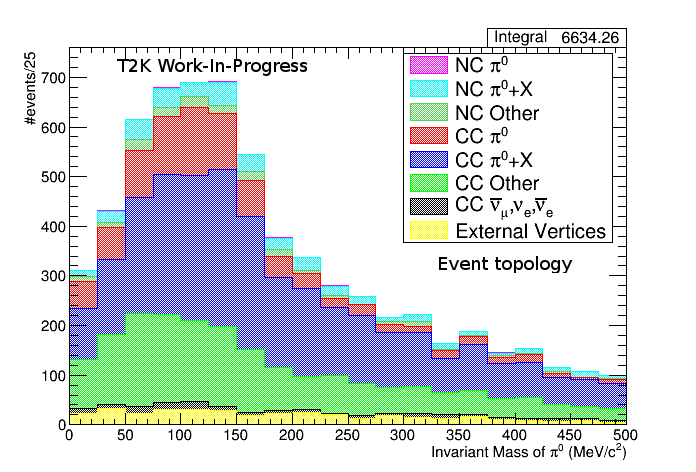}
\caption{LT emission  theta angle with respect to the neutrino direction for a signal sample (top-left), reconstructed LT momentum for a signal sample (top-right), reconstructed $\pi^{0}$ invariant mass for the selected signal sample (bottom-left), reconstructed $\pi^{0}$ invariant mass for a sideband sample (bottom-right)}
\label{fig:02}
\end{figure}

Using the above selection criteria, the candidate sample  of NC$\pi^{0}$ is selected. The purity selection is 24 $\%$ and the efficiency selection is 6 $\%$.
The main source of background is from CC interactions (Figure~\ref{fig:02} (bottom-left)).
Extra cuts used to reject $\mu$'s reconstructed only in the FGD or entering ECAL at large angles (Figure~\ref{fig:02} (top-left)) result in a minor reduction  in the background and a slight increase in the selection purity, but it drastically decreases the selection efficiency.
The use of $\pi^{+}$ for the selection of NC interactions increases the
 number of selected events (Figure~\ref{fig:02} (top-right)). 
A simultaneous fit to the selected NC$\pi^{0}$  
sample and the sideband sample can effectively  constrain the background. 
The sideband sample is obtained by reversing the muon veto cut in the signal selection. The idea behind the sideband sample is to select a sample with mostly a background that is  similar to the background of selected sample, NC$\pi^{0}$, and as small the signal as possible (Figure~\ref{fig:02} (bottom-right)). The simultaneous fit to the selected and sideband samples can help to estimate the number of signal and background events in the selected sample independently of MC simulations.
Initial studies have been performed and are being continued.

\section{Conclusion}

The discussion on how to improve the NC$\pi^{0}$ selection criteria and future plans were presented.
It is impossible to effectively eliminate the background in the NC$\pi^{0}$ sample using selection cuts only. 
The sideband sample should be used to estimate NC$\pi^{0}$  background.
The measurement of the inclusive NC$\pi^{0}$ production rate on CH can be done after background fitting is completed.

\Acknowledgements

This work was supported by the Polish National Science Centre grant 2015/17/D/ST2/03533,
UMO-2014/14/M/ST2/00850 and in part by PL-Grid Infrastructure.

\end{document}

%% file: econfmacros.tex
\def\beq{\begin{equation}}
\def\eeq#1{\label{#1}\end{equation}}
\def\eeqn{\end{equation}}

\def\beqa{\begin{eqnarray}}
\def\eeqa#1{\label{#1}\end{eqnarray}}
\def\eeqan{\end{eqnarray}}

\let\bar=\overbar

\def\Dslash{\not{\hbox{\kern-4pt $D$}}}
\def\dslash{\not{\hbox{\kern-2pt $\del$}}}

\def\msb{{\bar{\ssstyle M \kern -1pt S}}}

